\documentclass[prb, aps, twocolumn,superscriptaddress,showpacs]{revtex4}

\usepackage{graphicx}

\usepackage{amssymb, amsmath, color, rotating,bm}
\usepackage{color}

\newcommand{\br}{{\bf r}}
\newcommand{\bk}{{\bf k}}
\newcommand{\bq}{{\bf q}}
\newcommand{\bQ}{{\bf Q}}

\newcommand{\bS}{{\bf S}}

\begin{document}

\title{Interaction driven exotic quantum phases in spin-orbit coupled lattice spin$-1$ bosons}

\author{J. H. Pixley}
\email{jpixley@umd.edu}
\affiliation{Condensed Matter Theory Center and Joint Quantum Institute, Department of Physics, University of Maryland, College Park, Maryland 20742-4111 USA}
\author{Stefan S. Natu}
\affiliation{Condensed Matter Theory Center and Joint Quantum Institute, Department of Physics, University of Maryland, College Park, Maryland 20742-4111 USA}
\author{I. B. Spielman}
\affiliation{Joint Quantum Institute, National Institute of Standards and Technology,
and University of Maryland, Gaithersburg, Maryland, 20899, USA}
\author{S. Das Sarma}
\affiliation{Condensed Matter Theory Center and Joint Quantum Institute, Department of Physics, University of Maryland, College Park, Maryland 20742-4111 USA}

\begin{abstract}
We study the interplay between large-spin, spin-orbit coupling, and superfluidity for bosons in a two dimensional optical lattice, focusing on the spin-1 spin-orbit coupled system recently realized at the Joint Quantum Institute [Campbell et. al., arXiv:1501.05984].
 We find a rich quantum phase diagram, where, in addition to the conventional phases ---superfluid and insulator--- 
 contained in the spin-$1$ Bose-Hubbard model, 
 there are new lattice symmetry breaking phases. For weak interactions, the interplay between two length scales, the lattice momentum and the spin-orbit wave-vector induce a phase transition from a uniform superfluid to a phase where bosons simultaneously condense at the center and edge of the Brillouin zone at a non-zero spin-orbit strength. This state is characterized by spin density wave order, which arises from the spin-$1$ nature of the system. Interactions suppress spin density wave order, and favor a superfluid \textit{only} at the Brillouin zone edge. This state has spatially oscillating mean field order parameters, but a homogeneous density. We show that the spin density wave superfluid phase survives in a two dimensional harmonic trap, and thus establish that our results are directly applicable to experiments on $^{87}$Rb, $^7$Li,  and $^{41}$K. 
\end{abstract}

 
\maketitle

The interplay between spin-orbit coupling, lattice physics, and interactions is fundamental to many areas of condensed matter and materials physics  from topological insulators~\cite{Hassan-2010}, lattice analogues of quantum Hall effects~\cite{Haldane-1988}, spin liquids~\cite{Balents-2010}, and spintronics based devices~\cite{DasSarma-2004}.  Recent progress with ultra-cold atoms and molecules has allowed us to ``engineer" spin-orbit coupling and competing interactions in systems that have no solid state counterparts. This enables the study of fundamental questions such as the fate of superfluidity/superconductivity in the presence of single particle degeneracies 
\cite{Lin-2009,Lin-2011,Wang-2012, OurNature,Jo-2012,Aidelsburger-2011,Aidelsburger-2013,Kennedy-2015,Taie-2015}, the physics of bosons and fermions with large spin \cite{Ho-1998, Ohmi-1998},  quantum Hall effects in bosonic systems~\cite{Cooper2008}, and the physics of long range interactions \cite{Ni2008,Bendkowsky-2009,Lu-2011,Aikawa-2012}. In many cases, the very paradigms to qualitatively think about these strongly correlated systems is only now being developed.

\begin{figure}[h!]
\centering
  \includegraphics[width=1.0\linewidth]{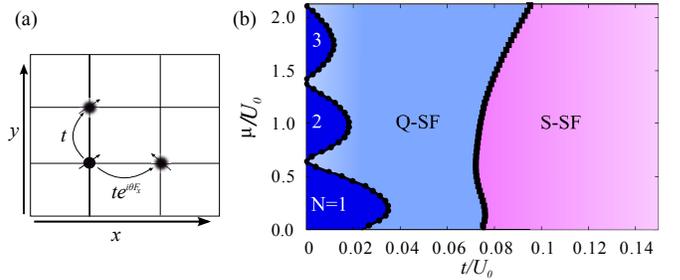}
\caption{(a) Schematic representation of the model on a square optical lattice.  Hopping in the $x$-direction is spin dependent carrying a SOC driven phase, while hopping in the $y$-direction is spin-independent with no phase. (b) Mean field phase diagram of the spin-$1$ spin orbit coupled Bose-Hubbard model
 with $U_2 = -0.25 U_0$, $\Omega_{R} = 0.1U_{0}$ and $\theta=\pi$. We find three distinct spin-orbit driven phases: a S-SF, a Q-SF, and density ($N$) tuned Mott phases at strong coupling (the phases are described in the text). Interactions suppress the SDW order 
 leading to a
 second order transition from the S-SF into the Q-SF phase, where the SF is now ferromagnetic and condensed at $\bq = (\pi,0)$. Complete translation symmetry is restored at a second transition to the ferromagnetic Mott phase. Our conclusions remain qualitatively valid even for weak $U_{2} = -0.005U_{0}$, which is the case for $^{87}$Rb.} 
 \label{fig:1}
\end{figure}

We theoretically study the interplay of spin-orbit coupling, large spin, and interactions by calculating the ground state phase diagram of a spin-$1$ spin-orbit coupled Bose gas in an optical lattice. 
In the continuum, spin-orbit coupling (SOC) gives rise to a single particle energy dispersion with multiple minima, 
each with a different spin character.  For weak interactions, and large on-site occupation number, bosons condense into multiple minima simultaneously, producing a ground state with a spatially oscillatory order parameter and density modulations~\cite{Zhai2010, Zhang2011, Wu2010,Li2012}, i.e. a stripe superfluid (SF). In the spin-$1$ system, the spin also develops a spatial texture through a spin density wave (SDW) SF~\cite{Cole2015, Lan2014, Ueda12}. This SDW SF has yet to be observed experimentally in any cold atom system.

As we show here, the physics in a lattice is strikingly different. In addition to inducing strong correlations by suppressing inter-particle tunneling, the lattice also suppresses any SF formation at incommensurate momentum. As a result, instead of a stripe superfluid modulating at the spin-orbit wave-vector for arbitrarily weak spin-orbit coupling, the lattice induces a second order phase transition between a homogeneous ferromagnetic (FM) SF to a stripe SDW superfluid (S-SF) at a strong, non-zero SOC strength. Remarkably, this stripe superfluid involves condensation at the center and the edge the Brillouin zone, and not at momenta (necessarily) commensurate with the spin-orbit wave vector. This phase transition is a direct result of the interplay between two length scales, namely the SOC wave-vector where the bosons prefer to condense in momentum space, and the lattice momentum, which describes where the bosons are pinned in real space.

Approaching the Mott limit simultaneously suppresses the superfluid and stripe order~\cite{Fisher1989, Jaksch1998}, and favors condensation at a single momentum, due to the 
competition of the various energy scales in the problem.  A key question then becomes 
whether SDW order and superfluidity are simultaneously destroyed at the Mott transition, or are these are two distinct quantum phase transitions? 
We find that the SDW superfluid is destroyed well before the Mott boundary at a continuous transition, giving rise to an interaction driven SF phase where condensation occurs at ${\bf q}=(\pi,0)$ on the Brillioun zone-edge. This novel superfluid has real space oscillations in the superfluid order parameter, but not in the density. Our predictions can be readily tested in experiments by generalizing the setup of Campbell \textit{et al.} \cite{Campbell2015} to a two dimensional optical lattice.

\emph{Spin-$1$ Bose-Hubbard model with SOC ---}
We consider the spin-$1$ Bose-Hubbard model~\cite{Imambekov-2003, Hickey2014} in two dimensions, in the presence of a one-dimensional spin-orbit coupling $H=H^t + \sum_i H^I_i$
\begin{eqnarray}
H^t&=& -t\sum_{\langle i,j \rangle}\left(a_{i\alpha}^{\dag}\mathcal{R}_{ij}^{\alpha\beta}a_{j\beta}+\mathrm{H.c} \right),
\\
H^I_i&=&  \frac{U_0}{2}n_i(n_i-1) + \frac{U_2}{2} \left(\bS^2_i-2n_i \right) +\Omega_R S_i^z+ V_i n_i,
\nonumber
\label{eqn:ham}
\end{eqnarray}
where $a_{i\alpha}^{\dag}$ describes the creation of a spin-$1$ boson in the $m_z=\alpha$ state,
repeated Greek indices are summed over, and $\langle i,j \rangle$ denotes a sum over nearest neighbors. 
We introduce the density $n_i =  a_{i\alpha}^{\dag}a_{i\alpha}$, the spin ${\bf S}_i = a_{i \alpha}^{\dag} {\bf F}^{\alpha \beta}a_{i\beta}$ (where $ {\bf F}$ is a vector of the spin-$1$ matrices), the spin-orbit coupling [via the spin dependent hopping see Fig.~\ref{fig:1} (a)] $\hat{\mathcal{R}}_{i,i+\hat{x}} =  e^{i \theta \hat{F}_x}$ and $\hat{\mathcal{R}}_{i,i+\hat{y}} =  \hat{1}$, the $3\times3$ identity matrix, the density-density interaction $U_0$, and the spin-spin interaction $U_2$. We study repulsive interactions $U_0 >0$, and a ferromagnetic spin-spin interaction $U_2<0$, as for $^{87}$Rb, $^7$Li,  and $^{41}$K (Ref.~\onlinecite{2013_SKurn_Ueda_RMP}). 
We work at a fixed chemical potential $\mu$, which enters the calculations through $V_i=W_i-\mu$, where $W_i$ is the harmonic trapping potential.
We include a Zeeman coupling (i.e. Raman coupling strength) $\Omega_R$ perpendicular to $S_x$, which is essential to ensure the SOC cannot be simply gauged away by a local unitary rotation. Importantly, the physics we discuss does not require strong spin-orbit coupling, and thus will not suffer from heating losses.

\begin{figure}[h!]
\centering
\begin{minipage}{.25\textwidth}
  \centering
  \includegraphics[width=0.9\linewidth]{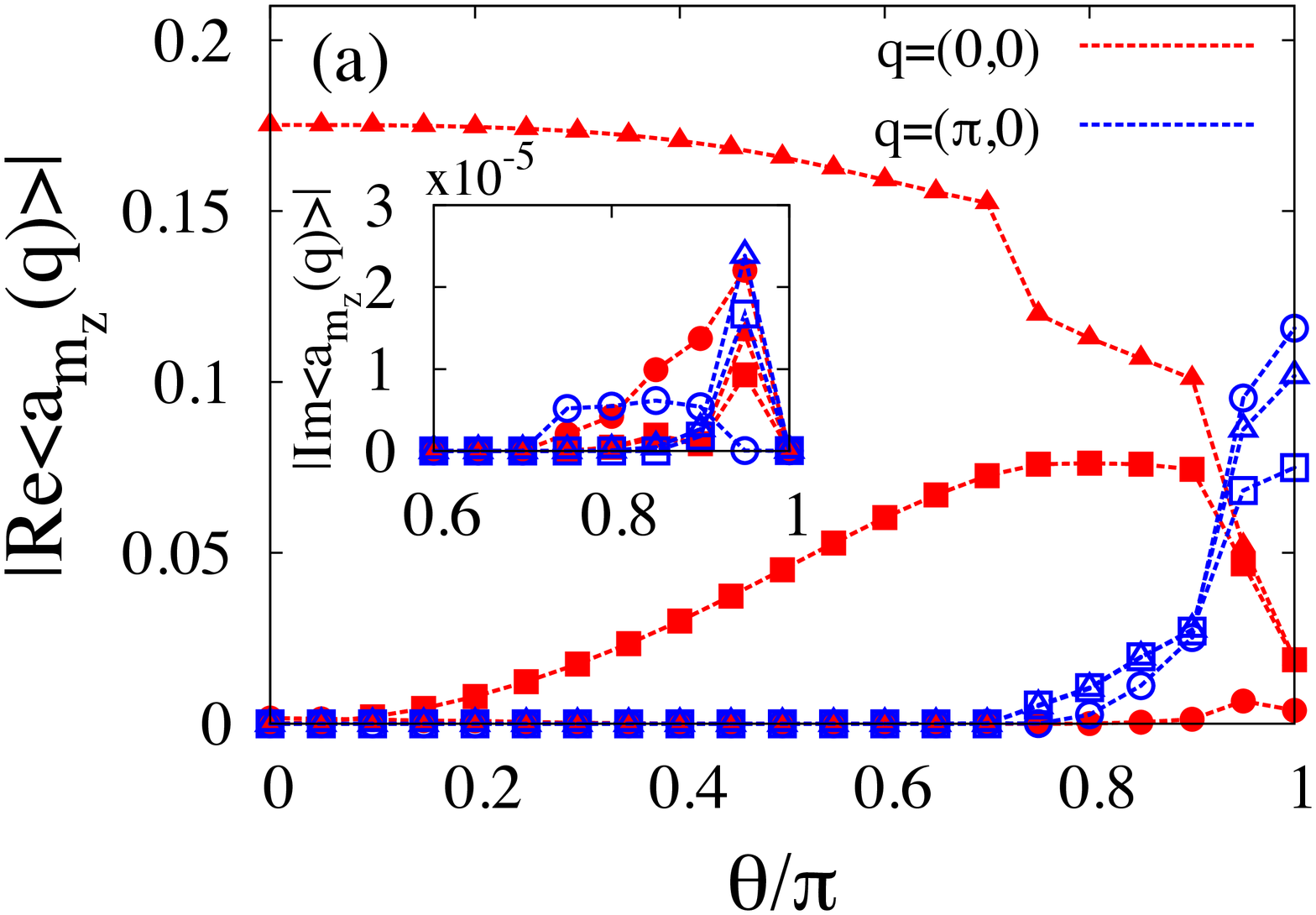}
\end{minipage}%
\begin{minipage}{.25\textwidth}
  \centering
  \includegraphics[width=0.9\linewidth]{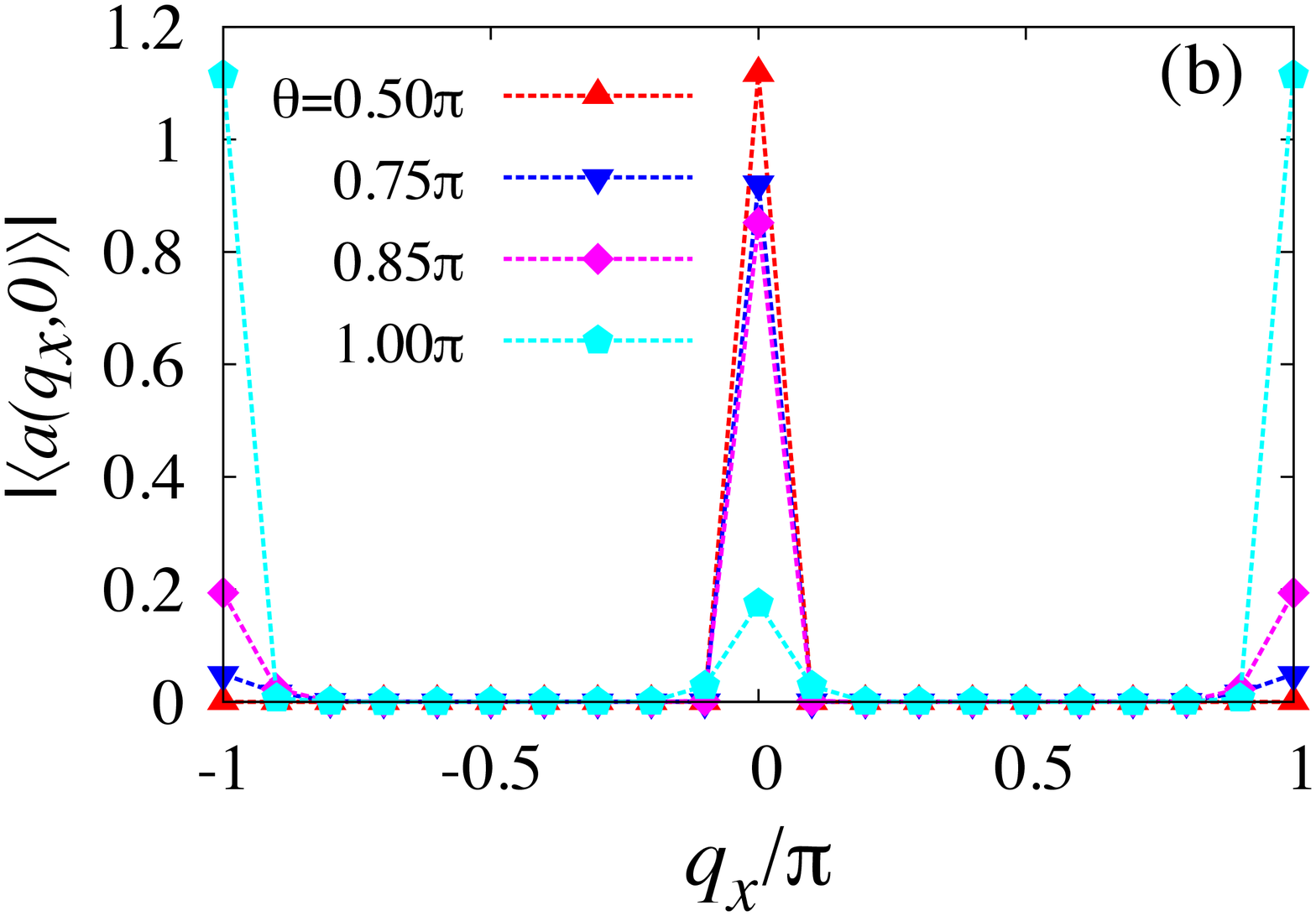}
 \end{minipage}
\begin{minipage}{.25\textwidth}
  \centering
  \includegraphics[width=0.7\linewidth,angle=-90]{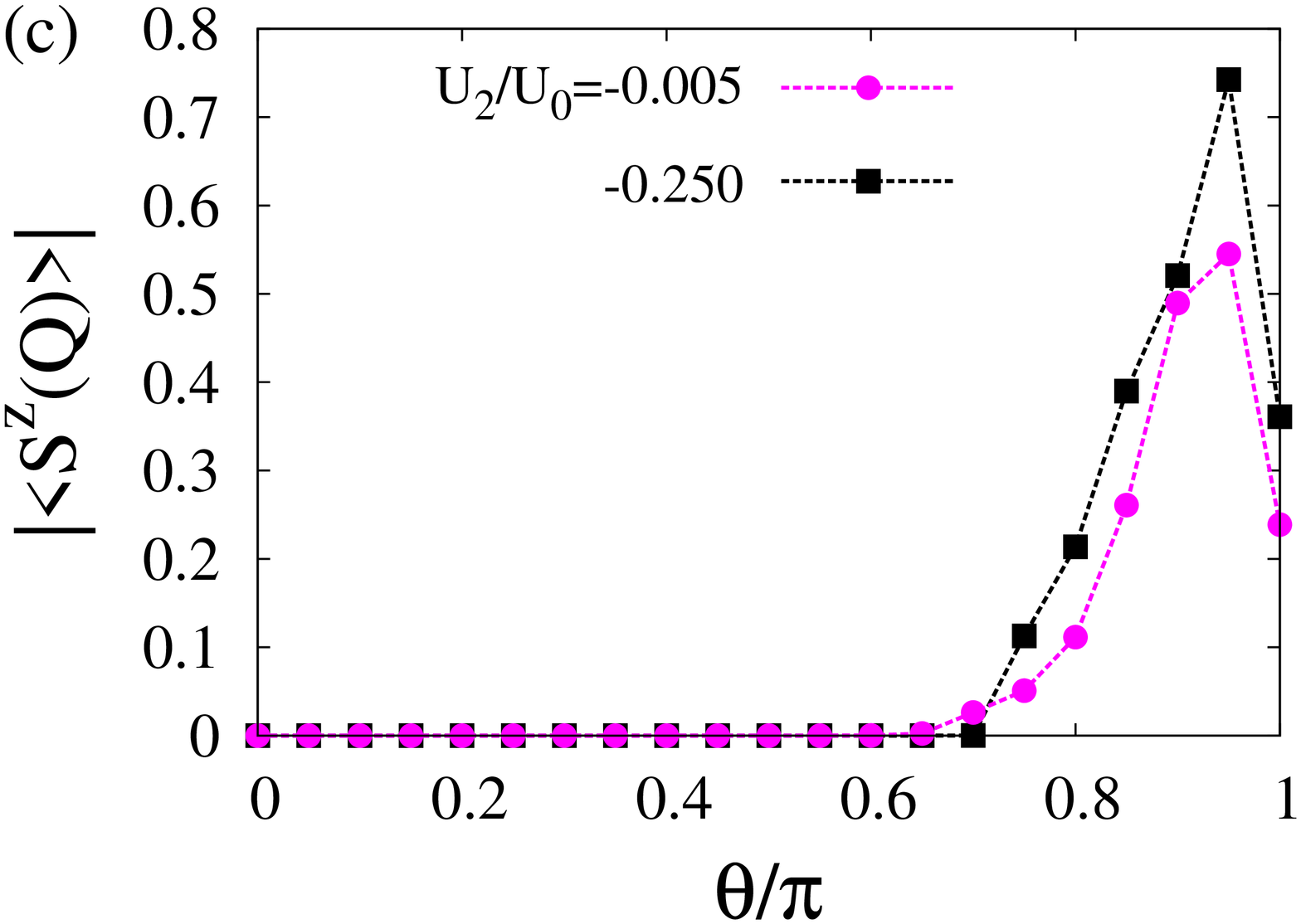}
\end{minipage}%
\begin{minipage}{.25\textwidth}
  \centering
  \includegraphics[height=0.8\linewidth,angle=-90]{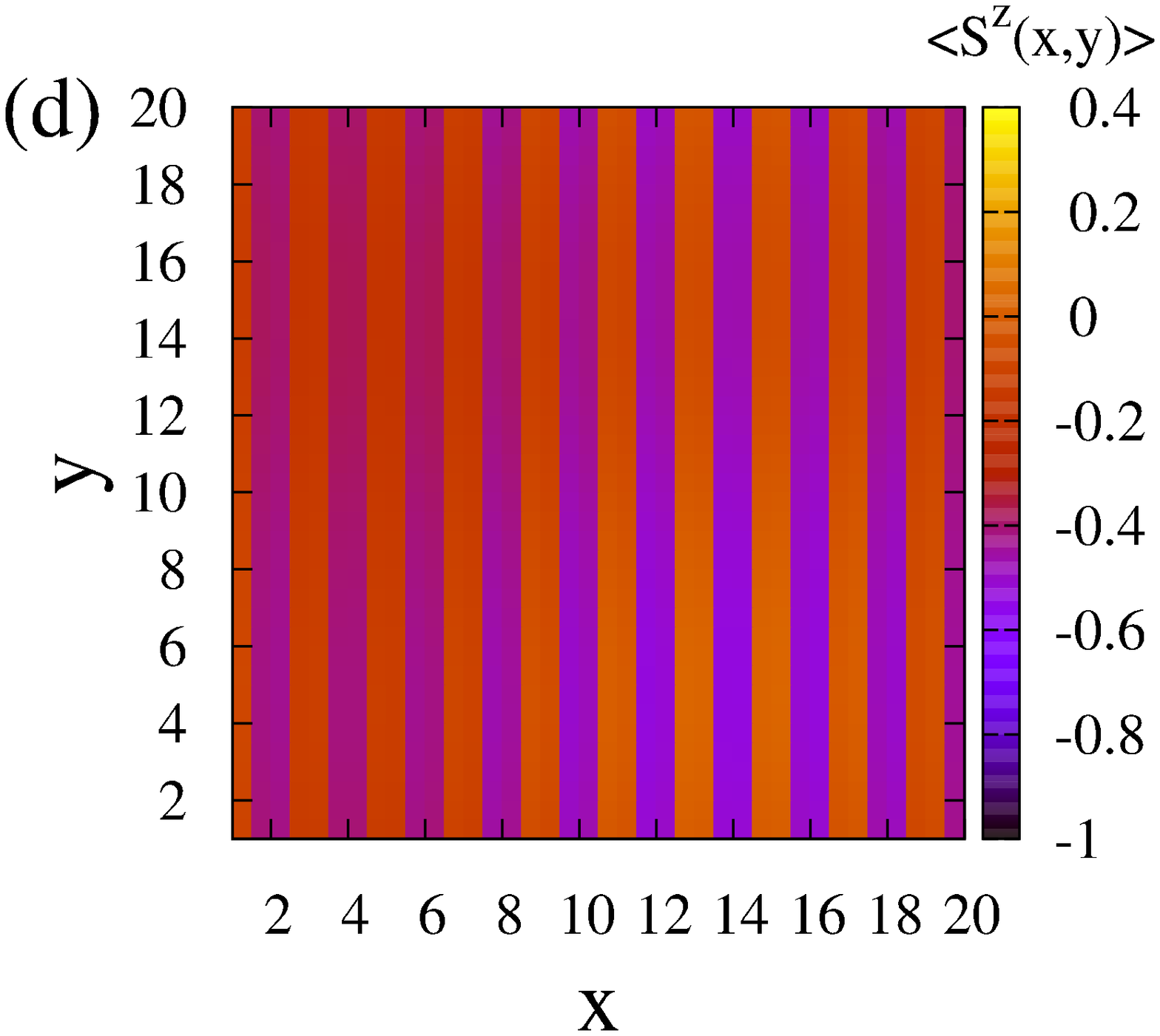}
\end{minipage}%
\caption{Evolution of the ferromagnetic superfluid as a function of the spin-orbit angle $\theta$ for $L_x=L_y=20$ for $\Omega_R=0.1U_0$, and $t = 0.2U_0$ [for panels (a), (b) and (d) $U_2=-0.25U_0]$. 
Absolute value of the real and imaginary (inset) parts of the superfluid order parameters for $m_z =1$ (squares), 0 (circles), and $-1$ (triangles)
at a momentum $\bq = (0,0)$ (red) and $\bq=(\pi,0)$ (blue) as a function of $\theta$. (b) The momentum dependence of $|\langle a(\bq) \rangle | \equiv \sqrt{|a_1(\bq)|^2+|a_0(\bq)|^2+|a_{-1}(\bq)|^2}$ for $q_y=0$, displaying that the weight of the mode at $\bq=(0,0)$  gets transferred to $\bq = (\pi,0)$ at the critical angle $\theta_c$. (c) The spin density wave (SDW) order parameter as a function of $\theta$ showing a continuous transition into the S-SF phase
 for $U_2/U_0 = -0.005$ and $-0.250$.  (d) The spin $\langle S^z(x,y) \rangle$ (in units of the average density) in real space, displaying a translationally invariant solution in the $y$-direction and a distinct SDW pattern.} 
 \label{fig:2}
\end{figure}

We use an inhomogenous Gutzwiller mean field theory (GMFT)~\cite{Rokshar-1991}, with a wavefunction \emph{ansatz} $|\psi  \rangle = \prod_{i} |\phi_i \rangle $. Where 
$| \phi_i \rangle = \sum_{m_1,m_0,m_{-1}} A_{i,(m_1,m_0,m_{-1})} |i;m_1,m_0,m_{-1}\rangle$
is the wave function at site $i$, and we determine the $A_{i,(m_1,m_0,m_{-1})}$ variationally, in an unbiased fashion allowing them to be different at each lattice site, while truncating each local Hamiltonian to a bosonic basis $N_b$. In the absence of SOC, the GMFT qualitatively describes polar and ferromagnetic~\cite{Ho-1998,Ohmi-1998} superfluid phases correctly~\cite{Kimura-2005,Pai-2008,Forges-2013,Natu-2015}, and also captures the presence of each Mott lobe. 
Here, we explore how this phase diagram is modified in the presence of the spin-orbit coupling, and in particular, how the superfluid and spin density wave order parameters evolve close to the Mott transition.  It is important to remark that quantum Monte Carlo \cite{Forges-2013}, which is often exact for bosonic systems incurs a sign problem from the complex hopping. Variational mean-field theories of the form we consider are thus essential to making progress.  

\begin{figure}[h!]
\centering
\begin{minipage}{.25\textwidth}
  \centering
  \includegraphics[width=0.88\linewidth]{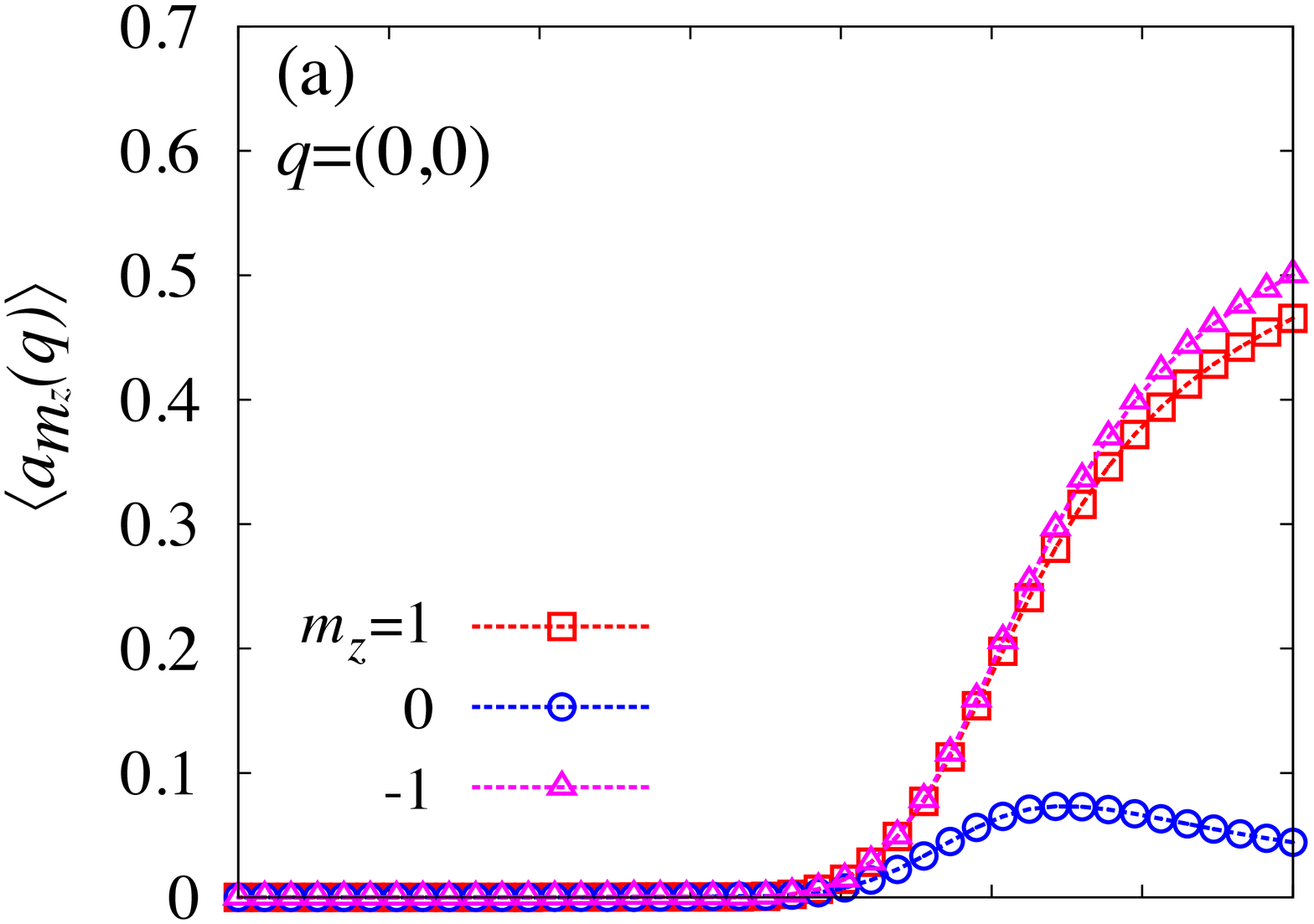}
\end{minipage}%
\begin{minipage}{.25\textwidth}
  \centering
  \includegraphics[width=0.85\linewidth]{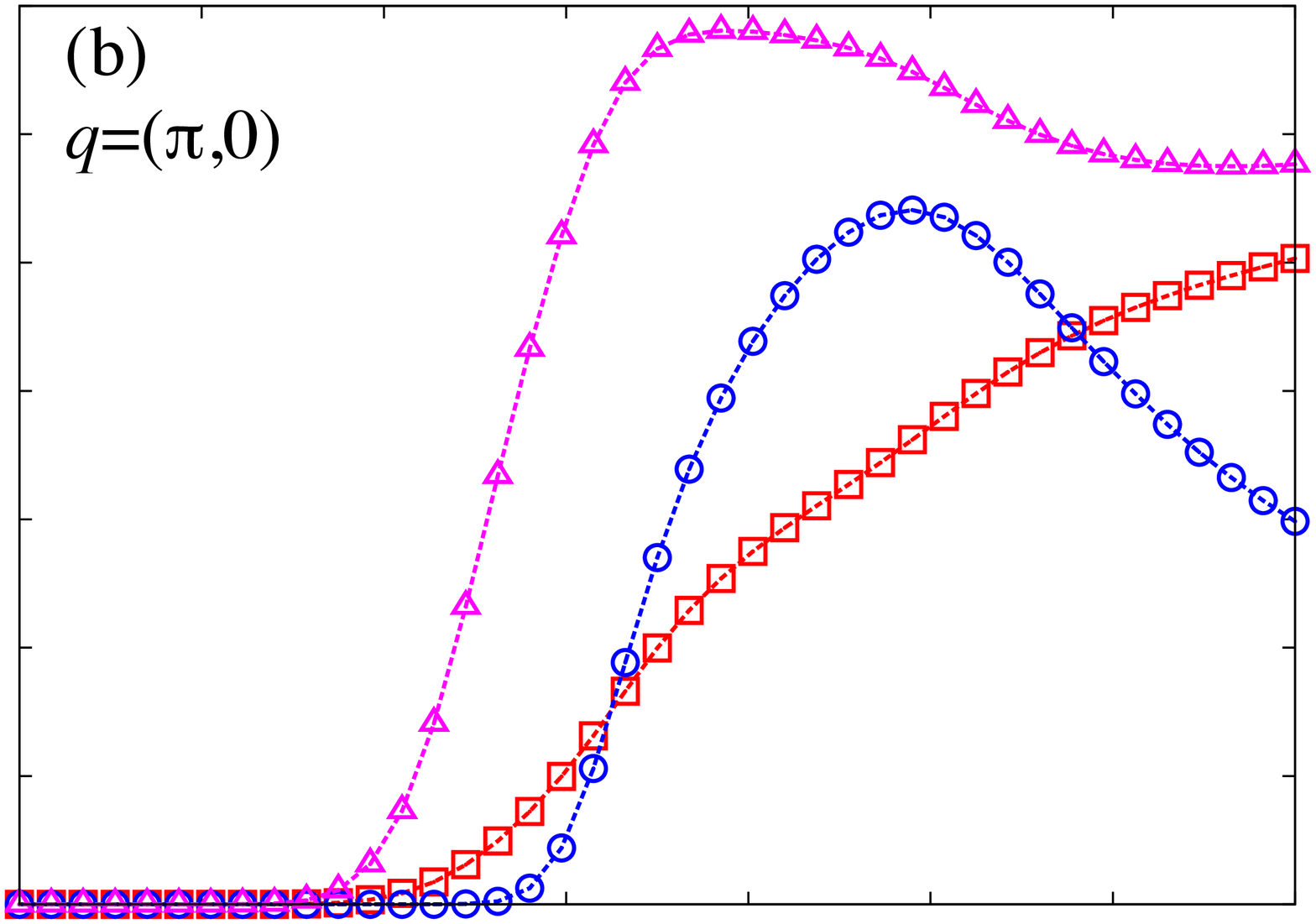}
 \end{minipage}
\begin{minipage}{.25\textwidth}
  \centering
  \includegraphics[width=0.95\linewidth]{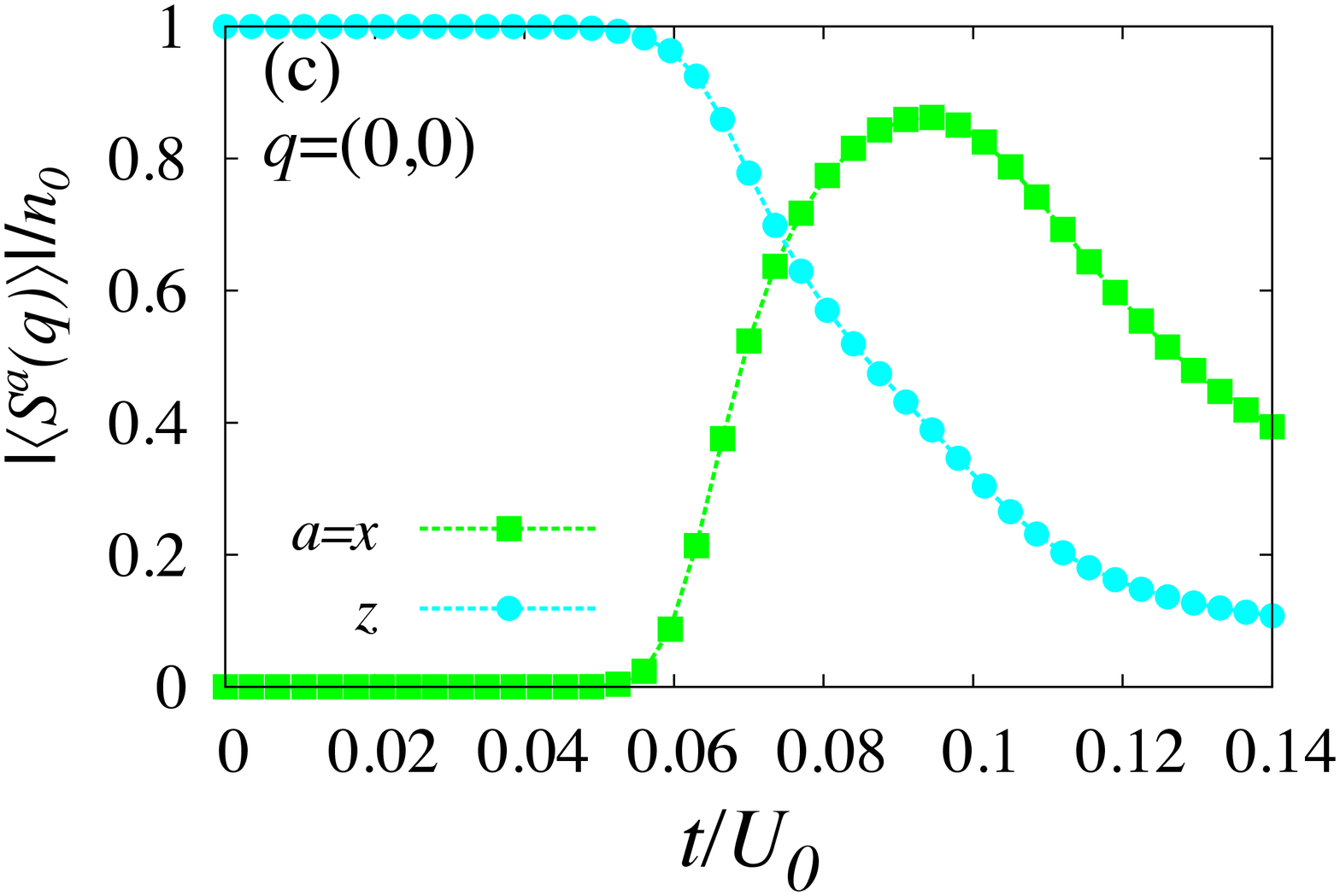}
\end{minipage}%
\begin{minipage}{.25\textwidth}
  \centering
  \includegraphics[width=0.88\linewidth]{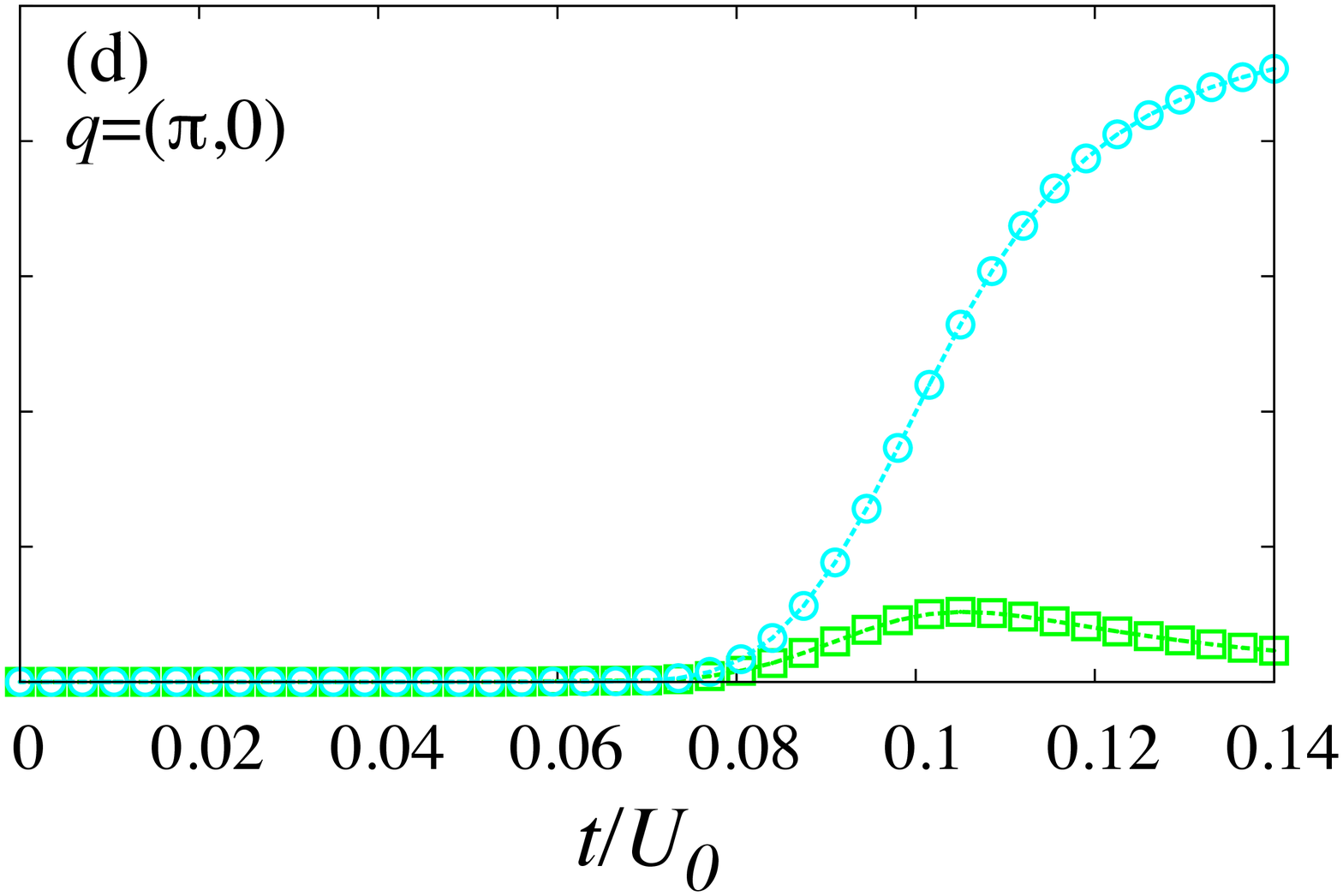}
\end{minipage}%
\caption{Evolution of the S-SF as a function of the strength of interaction $t/U_0$ for a fixed value of $\mu/U_0=0.23$, with translational invariance in the $y$-direction, $L_x=20$, $U_2=-0.25U_0$, $\Omega_R=0.1U_0$, $\theta=\pi$, and $N_b=4$.  The SF order parameter for each $m_z$ value at $\bq=(0,0)$ (a), and $\bq=(\pi,0)$ (b), with labels for $m_z$ shared between the two. The total spin (c) and the SDW order parameter (d) for both $x$ (green) and $z$ (cyan) components. The order parameters $\langle S^a(\bQ) \rangle$ and $\langle  a_{m_z}(0,0) \rangle$ are continuously suppressed at the S-SF to Q-SF transition [(a) and (d)], while the continuous Mott transition is captured by $\langle S^x \rangle$ and $\langle  a_{m_z}(\bQ) \rangle$ going to zero [(b) and (c)].} 
 \label{fig:3}
\end{figure}

We first calculate $\langle \psi | H |\psi  \rangle$, with no trap ($W_i=0$) and periodic boundary conditions (Figs~\ref{fig:1}, \ref{fig:2}, and \ref{fig:3}).
 The explicit form of $\langle \psi | \sum_i H_i^I |\psi  \rangle$, $\langle \psi | a_{i\alpha} |\psi  \rangle$, and $\langle \psi | a_{i\alpha}^{\dag} |\psi  \rangle$ can be found in Ref.~\onlinecite{Natu-2015}, while allowing the variational parameters to be site dependent.
  The hopping Hamiltonian becomes
$\langle \psi | H_t | \psi \rangle= -t\sum_{i,\alpha}\langle a_{i\alpha}^{\dag} \rangle R_{i\alpha}+\bar{R}_{i\alpha}\langle a_{i\alpha}  \rangle$,
where $R_{i\alpha} = \sum_{\langle j \rangle_i,\beta} \left[\mathcal{R}_{ij}^{\alpha\beta}+(\mathcal{R}_{ij}^{\alpha\beta})^*\right]\langle a_{j\beta} \rangle$, and 
$\bar{R}_{i\alpha} = \sum_{\langle j \rangle_i,\beta} \langle a_{j\beta}^{\dag} \rangle\left[\mathcal{R}_{ji}^{\beta\alpha}+(\mathcal{R}_{ji}^{\beta\alpha})^*\right]$, where $\sum_{\langle j \rangle_i}$ denotes a sum over the nearest neighbors of site $i$. 
We diagonalize each site of the resulting local Hamiltonian, which feels the neighboring condensate order parameter (through $R_{i\alpha}$) iteratively until the lowest energy state is reached.  To avoid getting trapped in local energy minima, we solve the mean field equations starting from different random initial conditions (for the $\langle a_{i\alpha} \rangle$), and find the solution with the lowest energy.
We consider small $\Omega_R = 0.1 U_0$, to avoid 
simply polarizing the spinor gas~\cite{Natu-2015}.
 In this work we focus primarily on the deep lattice limit ($ t \ll U_0$) and low particle number (considering bosonic Mott lobes up to $N=3$), and therefore can restrict the calculations to a small number of bosons per site.  As we show below, solving the fully two dimensional inhomogenous problem yields translationally invariant solutions in the $y$-direction, and a two site unit cell in the $x$-direction [see Fig.~\ref{fig:2}(d)]. Thus in constructing the phase diagram in Fig.~\ref{fig:1}, we consider translationally invariant solutions in the $y$-direction.

\emph{Striped Superfluid and Spin Density Wave Phase ---}
At weak coupling, the physics is dominated by the interplay of lattice effects and SOC, which we explore in Fig.~\ref{fig:2}.
We fix $U_2=-0.25 U_0$, $t = 0.2 U_0$ for lattice sizes $L_x=L_y=20$ (setting the lattice spacing to unity), and choose a small value of the chemical potential $\mu=0.5 U_0$, such that a $N_b=3$ basis is sufficient on each site.
We remark that this value of $U_2$ is experimentally realistic as $^7$Li has $U_2/U_0 \approx -0.5$ (Ref.~\onlinecite{2013_SKurn_Ueda_RMP}). 
 We take the Fourier transform of the bosonic operators to determine their structure in momentum space via $a_{\alpha}(\bq)= L_x^{-1}L_y^{-1}\sum_{\br }e^{i\bq\cdot\br_i}a_{i,\alpha}$. For $\theta=0$, the ground state is a ferromagnetic superfluid with $\langle a_{-1}(\bq =0) \rangle \neq 0$ (where the term containing $\Omega_R$ in Eq.~(\ref{eqn:ham}) selected the ordering direction), as shown in Fig.~\ref{fig:2}(a) and Refs.~\cite{Ho-1998,Ohmi-1998,Kimura-2005,Pai-2008,Forges-2013,Natu-2015}. 
A spiral magnetic state develops with increasing $\theta$ [Fig.~\ref{fig:2} (a)]
as $\langle a_{1}(\bq =0) \rangle \neq 0 $ for small $\theta$. 

As $\theta$ increases further, we find a continuous quantum phase transition at $\theta_c(t,U_2=-0.25U_0,\Omega_R)\approx 2.2$ into a SDW superfluid phase (S-SF). We confirmed that increasing $N_b$ from $3$ to $5$ leaves the physics unchanged. This transition is independent of a smaller value of $U_2$ closer to that of $^{87}$Rb, see Fig~\ref{fig:2}(c).
For $\theta > \theta_c$, the condensate density at the point $(\pi, 0)$ on the edge of the Brillouin zone rises continuously as $\theta$ increases further [Fig.~\ref{fig:2} (a) and (b)].
Here, we observe the condensate density rising continuously at $\langle a_{m_z}(\bQ) \rangle$, where $\bQ \equiv (\pi,0)$, for each $m_z$ component [Fig.~\ref{fig:2}(a)], and acquires a non-zero imaginary part 
in both momentum components ($\bq =0, \bQ$), 
[inset of Fig.~\ref{fig:2} (a)].  The precise location of this phase transition is determined by an interplay between the single particle spin-orbit physics and the underlying lattice, as we explain below. The non-zero momentum condensate is pinned to an ordering vector $\bQ_{\mathrm{or}}$ commensurate with the underlying lattice $\bQ_{\mathrm{or}}=\bQ $ for $\theta_c \leq \theta \leq \pi$, as shown in Fig.~\ref{fig:2}(b). No other momentum components condense. 

In the continuum, the SDW phase results purely from non-interacting bosons condensing into multiple minima of the single-particle dispersion \cite{Cole2015, Lan2014}. Here, the lattice suppresses condensation into incommensurate spin-orbit minima, until the spin-orbit wave vector is large enough such that the single-particle minima becomes close to $(\pi, 0)$. This is because in the limit of small on-site occupation, it costs a large amount of energy to condense away from a lattice site, as a result the superfluid is pinned to momenta commensurate with the lattice wave-vector rather than the spin-orbit wave-vector. We have checked that our calculations reproduce continuum results at large on-site occupation ($N_{\text{b}} \sim 10$) and $t/U_0$, where the lattice no longer plays a dominant role in the physics.

A second non-trivial feature of the spin-$1$ system (compared to its pseudo spin-$1/2$ counterpart) is the appearance of a longitudinal SDW, concomitant with density wave order \footnote{We caution that SDW order is indeed possible in spin-$1/2$ systems with Rashba dispersion \cite{Yu-2013},  but this is significantly more challenging to realize in cold gases. Furthermore, spin density wave order is present in the transverse components, but this typically follows the spin-orbit field, which explicitly breaks spin symmetry.}.
We take 
$
\langle S^a(\bQ) \rangle
$
 as the SDW order parameter, which rises continuously upon entering the S-SF phase as shown in Fig.~\ref{fig:2} (c).  
 We find that the SDW is well described by $\langle S^a(x,y) \rangle = \mathcal{A}_{S^a}+\mathcal{B}_{S^a} \cos(\pi x)$,  where $a=x,y,z$ and an amplitude $\mathcal{B}_{S^a}$ on the order of the condensate density [see Fig.~\ref{fig:2} (d)].   
 
 To understand the origin of the longitudinal SDW, note that deep in the S-SF phase we can treat the spin order parameter at the mean field level, i.e. $\langle {\bf S}(\bq) \rangle \approx \sum_{\bk} \langle a_{\alpha}(\bq + \bk)^{\dag} \rangle {\bf F}_{\alpha \beta} \langle a_{ \beta}(\bk) \rangle$. Therefore, the amplitude and wave-vector of the spin density wave order is completely tied to the existence of a condensate at $\bq = (0,0)$ and $(\pi,0)$.  We also find a very weak charge density wave (not shown), however it is strongly penalized by the lattice and interactions. 
A major advantage of the spin-$1$ system over the spin-$1/2$ counterpart is that the existence of additional longitudinal SDW order should make the stripe superfluid phase easier to detect experimentally. In the spin-$1/2$ case with NIST type SOC, there is no longitudinal SDW at zero detuning, and the contrast of the density oscillations is typically extremely small \cite{Li2012, Zhang2011, Wu2010, Zhai2010}.

\emph{Strong Coupling Phases ---}
Having established the weak coupling physics and the nature of  the S-SF phase, we turn to the central question of this paper: how do strong interactions compete/complement the spin-orbit coupling to affect the condensate and spin/charge density wave orders? To address this, we now explore the phase diagram near the Mott-superfluid boundary at fixed SOC wave-vector $\theta=\pi$, and tune the ratios $\mu/U_0$ and $t/U_0$, which will eventually drive the model to a Mott insulating state \cite{Kimura-2005,Pai-2008,Forges-2013,Natu-2015}.

Fig.~\ref{fig:1} shows the resulting phase diagram at $\theta=\pi$.
We present results for $L_x=20$ (with translational symmetry along $y$) with $N_b=4$ and $U_2/U_0=-0.25$.
For large $t/U_0$ we find the S-SF phase, which is well described by our previous discussion for a large region of the parameter space. 
As shown in Figs.~\ref{fig:3} (a) and (d), 
decreasing $t/U_0$ sends $\langle a_{m_z}(\bq=0) \rangle \rightarrow 0$, which destroys the SDW order continuously $\langle S^a(\bQ) \rangle \rightarrow 0$, giving way to a superfluid where $\langle a_{m_z}(\bQ) \rangle  \neq 0$ and $\langle S^a(\bq=0) \rangle\neq 0$ ($a=x,z$) [see Figs.~\ref{fig:3} (b) and (c)], 
 which we  
define as
the Q-SF phase due to the condensate only appearing at a finite momentum. Increasing $\Omega_R$ 
first shifts the S-SF/Q-SF phase boundary to larger $t$, and finally completely polarizes the gas. Decreasing $\theta$, shrinks the regime of stability of the Q-SF phase, nonetheless a $\theta \approx \pi$ is experimentally realistic.

The phase transition from the S-SF to the Q-SF phase is a non-trivial interplay between competing energy scales in the problem.
Upon decreasing $t/U_0$, interactions suppress the overall amplitude of the condensate, and also favor condensation at a single momentum,  because of the additional exchange energy between condensed bosons at $\bq = (0, 0)$ and $(\pi, 0)$. In the continuum, local interactions induce a phase transition to a ``plane wave state" \cite{Li2012}, where bosons either condense at one or the other single-particle minima, a \textit{spontaneously} broken symmetry. In a lattice however, a state with a spatially modulating order parameter ($\langle a_{m_z}(\bQ) \rangle  \neq 0$), but a homogeneous density has lower spin-orbit \textit{kinetic} energy than a state with a uniform order parameter ($\langle a_{m_z}(\bq = 0) \rangle  \neq 0$) and density. 
This can be seen for $\theta=\pi$ by applying a spin rotation about the $y$-axis ($S_x \rightarrow -S_z$ and $S_z \rightarrow S_x$), which results in the hopping matrix element switching sign $t \rightarrow -t$ for $m_z = \pm 1$, favoring condensation at finite, rather than zero momentum \cite{Dhar-2013}. 
The Q-SF phase, where the superfluid order parameter has the same amplitude, but switches sign from site to site, minimizes the kinetic, interaction and Zeeman energies. This phase is therefore \textit{universally} picked as the lower energy state. We remark that such finite momentum condensed phases of bosons are rather non-generic, and are actively being studied for bosons in ladder geometries with artificial magnetic fields~\cite{Dhar-2013}.

As shown in Figs.~\ref{fig:3}(b) and (c), as 
$t/U_0$ further decreases $\langle S_x \rangle, \langle a_{m_z}(\bQ) \rangle$ vanishes continuously at the Mott transition, see Fig.~\ref{fig:3}(b) and (d). Putting all of this together we arrive at our main results shown in Fig.~\ref{fig:1}. 
Importantly, the transition into each ferromagnetic Mott lobe is distinct from the case $\theta=0$ (absence of SOC), since it is now the Q-SF phase which is becoming gapped out. Within the GMFT, each Mott lobe is homogeneous and ferromagnetic, as the Mott state is only captured at the $t=0$ level and particle-hole and spin fluctuations are neglected.

\begin{figure}[h]
\begin{minipage}[c]{.25\textwidth}
\hspace*{-4mm}
  \includegraphics[width=0.8\linewidth,angle=-90]{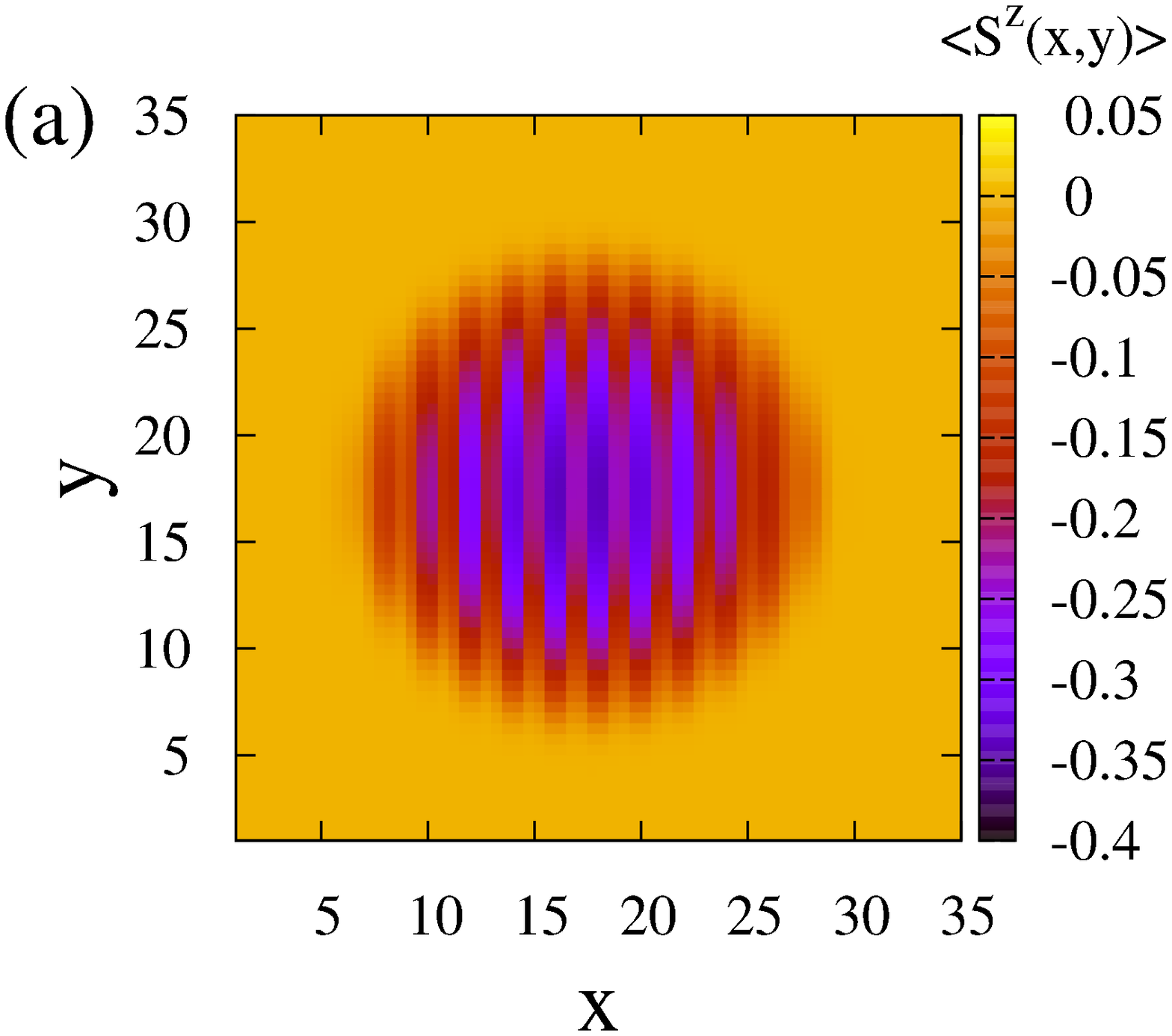}
\end{minipage}%
%
\hspace*{-4mm}
\begin{minipage}[c]{.25\textwidth}
 \includegraphics[width=0.8\linewidth,angle=-90]{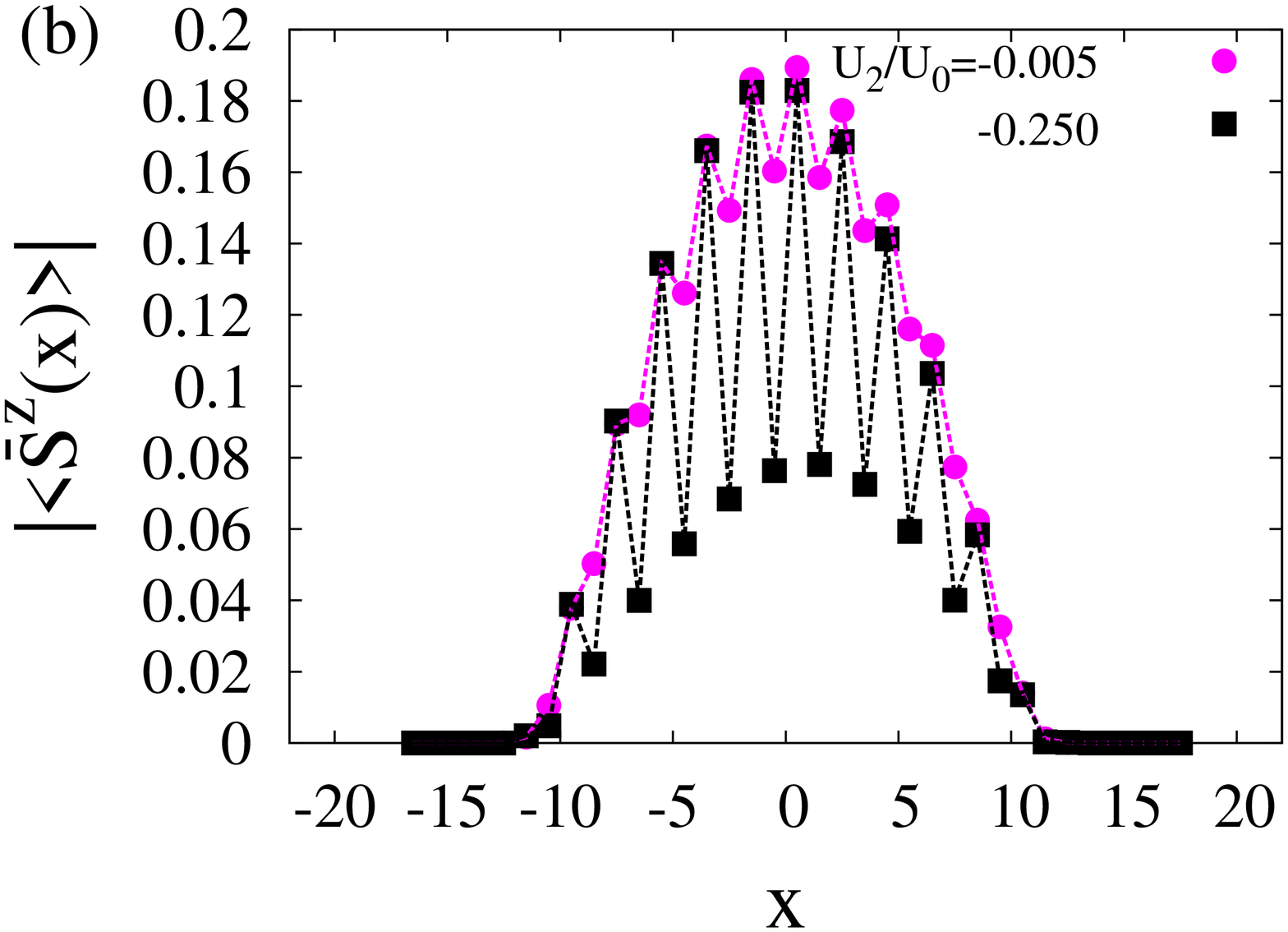}
 \end{minipage}
 \caption{ Spin density wave of the S-SF phase in a harmonic trap with $t=0.2 U_0$, $\mu = 1.0 U_0$, $\lambda=0.01 U_0$, $\Omega_R=0.1U_0$, $L_x=L_y=35$, $\theta=\pi$, and $N_b=5$. (a) SDW for $U_2/U_0=-0.25$, the color displays the value of $|\langle S^z(x,y) \rangle|$ in the $x-y$ plane. (b) Integrated spin density $\langle \bar{S}^z(x)\rangle\equiv \sum_y\langle S^z(x,y)\rangle/L$ as a function of $x$ for $U_2/U_0=-0.005$ and $-0.25$.} 
 \label{fig:4}
\end{figure}

The excitation spectrum of the superfluid phases we find will be quite interesting. In the Q-SF, we expect a quadratically dispersing ferromagnetic spin mode about $\bq = (0,0)$ and a density mode which is linearly dispersing at small $|\bq|$ about $\bQ$, but has a roton-maxon like structure at intermediate $q$. This structure arises because bosons condense into only one of the single particle minima, effectively gapping out the other. The transition from the Q-SF to the S-SF should be marked by the vanishing of the roton gap.  A detailed calculation of the excitation spectrum in the spin-$1$ SOC Bose gas will be the subject of future work. 

\emph{Experimental Relevance ---}
Our system can be readily studied in experiments by generalizing the existing setup of Campbell \textit{et al.} to an optical lattice \cite{Campbell2015}. Experimentally, the detection of $\langle S^z \rangle$ is much more challenging than $\langle S^x \rangle$, while a precise measurement of each $\langle a_{m_z} \rangle$ is possible.
Within the experimental apparatus of the NIST setup \cite{Campbell2015}, and the measurement basis we choose, the spin density wave in $ S^x  $ [Figs.~\ref{fig:3} (c) and (d)] should be experimentally accessible. We have also numerically confirmed the existence of the S-SF phase for $U_2=-0.005 U_0$ applicable to $^{87}$Rb [see Figs.~\ref{fig:2}(c) and ~\ref{fig:4}(b)].   Our results in Figs.~\ref{fig:1}-\ref{fig:3} provide a theoretical platform to interpret and understand experimental data. A key ingredient in most experiments is a harmonic trap  $W_i = \lambda (x_i^2+y_i^2)$ (we assume strong confinement in the $z$-direction, and use open boundary conditions). In Fig.~\ref{fig:4}(a) we plot the $\langle S^z(x,y) \rangle$ order parameter for $U_2/U_0=-0.25$ taking into account the external trapping profile, which shows oscillations that would be measurable \textit{in situ}. Fig.~\ref{fig:4}(b) shows the dependence of the magnitude of spin modulations within the trap on $U_2$, where for a small value (applicable to $^{87}$Rb) the SDW has non-zero yet small modulations, which increase dramatically for $U_2=-0.25U_0$ (closer to that of $^7$Li). Both the Q-SF and the S-SF can also be readily studied in time-of-flight, which should reveal a single peak for the former and  a bimodal structure for the latter. Furthermore, the different excitation spectra in the two cases is another experimental probe, which can be measured using Bragg spectroscopy \cite{Stenger99}. Our studies highlight the non-trivial interplay between large spin, lattice physics and interactions and will serve to guide ongoing experiments on these systems.

\acknowledgements{\emph{Acknowledgements ---} We thank W. Cole, E. Mueller, S. Chakram, M. Vengalattore and D. Trypogeorgos for useful discussions. We are grateful for the support provided by LPS-CMTC, LPS-MPO-CMTC, 
AROÕs Atomtronics MURI, the AFOSRÕs Quantum Matter MURI, NIST, the NSF through the PFC at the JQI,
 and the PFC seed grant ``Emergent phenomena in interacting spin-orbit coupled gases" (JHP, SN) for support. }

\bibliography{s1BH-SOC}

\end{document}